\documentclass[aps,pre,superscriptaddress]{revtex4}
\pdfoutput=1
\usepackage[a4paper, margin=3cm]{geometry}
\usepackage{graphicx}
\usepackage{bm}

\usepackage{tablefootnote}
\usepackage{amsmath}
\usepackage{amssymb}
\usepackage{url}
\newcommand{\e}[1]{\ensuremath{\times 10^{#1}}}
\newcommand{\ourgamma}{1.15650(50)}
\newcommand{\ournu}{0.58785(40)}
\newcommand{\ourmufcc}{10.037075(20)}
\newcommand{\ourmubcc}{6.530520(20)}
\newcommand{\ourAfcc}{1.1736(24)}
\newcommand{\ourAbcc}{1.1785(40)}
\newcommand{\ourDfcc}{1.0460(50)}
\newcommand{\ourDbcc}{1.0864(50)}

\newcommand{\mubcc}{\mu_{\rm bcc}}
\newcommand{\Abcc}{A_{\rm bcc}}
\newcommand{\Dbcc}{D_{\rm bcc}}
\newcommand{\mufcc}{\mu_{\rm fcc}}
\newcommand{\Afcc}{A_{\rm fcc}}
\newcommand{\Dfcc}{D_{\rm fcc}}

\usepackage[hidelinks]{hyperref}

\hypersetup{
    pdfauthor={Raoul D. Schram, Gerard T. Barkema, Rob H. Bisseling, Nathan Clisby},
    pdftitle={Exact enumeration of self-avoiding walks on BCC and FCC lattices}
    }
\begin{document}

\title{Exact enumeration of self-avoiding walks on BCC and FCC lattices}
\author{Raoul D. Schram}
\affiliation{Laboratoire de Physique,
Ecole Normale Sup\'{e}rieure de Lyon,
46, all\'{e}e d'Italie, 69364 Lyon, Cedex 07, France}
\affiliation{Mathematical Institute, Utrecht University,
P.O. Box 80010, 3508 TA Utrecht, The Netherlands}
\author{Gerard T. Barkema}
\affiliation{Department of Information and Computing Sciences, Utrecht University,
P.O. Box 80089, 3508 TB Utrecht, The Netherlands}
\author{Rob H. Bisseling}
\affiliation{Mathematical Institute, Utrecht University,
P.O. Box 80010, 3508 TA Utrecht, The Netherlands}
\author{Nathan Clisby}
\affiliation{School of Mathematics and Statistics, University of Melbourne, Australia}

\date{March 27, 2017}

\begin{abstract}
    Self-avoiding walks on the body-centered-cubic (BCC) and
    face-centered-cubic (FCC) lattices are enumerated up to lengths 28
    and 24, respectively, using the length-doubling method.  Analysis of
    the enumeration results yields values for the exponents $\gamma$ and
    $\nu$ which are in agreement with, but less accurate than those
    obtained earlier from enumeration results on the simple cubic
    lattice. The non-universal growth constant and amplitudes are
    accurately determined, yielding for the BCC lattice $\mu=\ourmubcc$,
    $A=\ourAbcc$, and $D=\ourDbcc$, and for the FCC lattice
    $\mu=\ourmufcc$, $A=\ourAfcc$, and $D=\ourDfcc$.
\end{abstract}

\maketitle

\section{Introduction}

The enumeration of self-avoiding walks (SAWs) on regular lattices
is a classical combinatorial problem in statistical physics, with a
long history, see e.g.~\cite{madras93,janse15}. 
Of the three-dimensional lattices, the simple cubic (SC) lattice has
drawn the most effort, starting with a paper by Orr~\cite{orr47}
from 1947, where the number of SAWs $Z_N$ was given for all $N$
up to $N_{\max}=6$; these results were obtained by hand. In 1959,
Fisher and Sykes~\cite{fisher59} used a computer to enumerate all SAWs
up to $N_{\max}=9$; Sykes and collaborators extended this to
11 terms in 1961~\cite{Sykes1961SomeCountingTheorems}, 16 terms in
1963~\cite{Sykes1963SelfAvoidingWalksontheSimpleCubicLattice}, and 19
terms in 1972~\cite{Sykes1972SAWsAndSARsOnVariousLattices}. 
In the following decade
Guttmann~\cite{guttmann87} enumerated SAWs up to $N_{\max}=20$ in 1987,
and extended this by one step in 1989~\cite{guttmann89}. In 1992,
MacDonald et al.~\cite{macdonald92} reached $N_{\max}=23$, and in 2000
MacDonald et al.~\cite{macdonald00} reached $N_{\max}=26$.  In 2007,
a combination of the lace expansion and the two-step method allowed for
the enumeration
of SAWs up to $N_{\max}=30$ steps~\cite{clisby07}.
Recently, the length-doubling method~\cite{schram11} was presented
which allowed enumerations to be extended up to $N_{\max}=36$.
To date, this is
the record series for the SC lattice. 

The body-centered-cubic (BCC)
and face-centered-cubic (FCC) lattices are in principle equally as
physically relevant as the SC lattice, but 
enumeration is hampered
by the larger lattice coordination numbers, which detriments
most enumeration methods severely. It is also very slightly more
cumbersome to write computer programs to perform enumerations for these
lattices. Consequentially, the SC lattice has served as the
test-bed problem for new enumeration algorithms, and the literature on
enumerations for the BCC and FCC lattices is far more sparse.
For the BCC lattice, $Z_N$ was determined up to 
$N_{\max}=15$ in 1972~\cite{Sykes1972SAWsAndSARsOnVariousLattices}, 
and to $N_{\max}=16$ in 1989~\cite{guttmann89}.
The current record of $N_{\max}=21$ was obtained in
1997 by Butera and Comi~\cite{Butera1997Nvectorspin} as the $N \rightarrow 0$ limit of the
high temperature series for the susceptibility of the $N$-vector model. 
For the FCC lattice, 
enumerations up to $N_{\max}=12$
were performed in
1967~\cite{Martin1967ProbabilityofInitialRingClosureGorSelfAvoidingWalks},
and the record of
$N_{\max}=14$ was achieved way back in
1979~\cite{McKenzie1979SAWsFCC}.

Enumeration results derive their relevance from the ability to 
determine critical exponents, which, according to renormalization group theory,
are believed to be shared between SAWs on various lattices and real-life
polymers in solution~\cite{degennes79}. Two such exponents are the entropic
exponent $\gamma$ and the size exponent $\nu$.
Given the number $Z_N$ of SAWs of all lengths up to $N_{\max}$
and the sum $P_N$ of their squared end-to-end extensions, these two
exponents can be extracted using the relations
\begin{align}
    Z_N &= A \mu^N N^{\gamma-1} \left(1 + \frac{a}{N^{\Delta_1}} +
    O\left(\frac{1}{N}\right)\right); \label{eq:zn} \\
    \frac{P_N}{Z_N} &= \sigma D N^{2\nu} \left(1 + \frac{b}{N^{\Delta_1}} +
    O\left(\frac{1}{N}\right)\right). \label{eq:rhon}
\end{align}
In these expressions, the growth constant $\mu$ and the amplitudes
$A$ and $D$ are non-universal (model-dependent) quantities, while the
leading correction-to-scaling exponent is a universal quantity with
value $\Delta_1 =
0.528(8)$~\cite{Clisby2016HydrodynamicRadiusForSAWs}. 
Sub-leading corrections-to-scaling are absorbed into the $O(1/N)$ term.
$\sigma$ is a lattice specific constant to ensure that our amplitude
``$D$'' is the same as in earlier work. $\sigma$ corrects for the fact that
with our definition each step of the walk is of length $\sqrt{2}$
for the BCC lattice (leading to $\sigma = 2$), and of length $\sqrt{3}$
for the FCC lattice (leading to $\sigma = 3$).
Note that for bipartite lattices, of which the SC and BCC lattices are
examples,
there is an additional alternating ``anti-ferromagnetic'' singularity,
that is sub-leading but which still must be treated carefully as the
odd-even oscillations tend to become amplified by series analysis
techniques.
Because of
universality, the exponents are clearly more interesting from a physics
perspective.  However, accurate estimates for the growth constant
and the amplitudes can also be very helpful for many kinds of computer
simulations on lattice polymers.

In this paper, we used the length-doubling method~\cite{schram11} to
measure $Z_N$ and $P_N$ up to $N_{\max}=28$ and 24, on the BCC and FCC
lattices, respectively. These lattices can be easily simulated as
subsets of the SC lattice: the collection of sites in which $x$, $y$ and
$z$ are either all even or all odd forms a BCC lattice, and the
collection of sites $(x,y,z)$ constrained to even values for $x+y+z$
forms a FCC lattice.  We then analysed these results to obtain estimates
for the exponents $\gamma$ and $\nu$, the growth constant $\mu$,
and the amplitudes $A$ and $D$. Our results for the two exponents
$\gamma$ and $\nu$ agree with the values reported in literature which
are obtained on the SC lattice, reinforcing the credibility of the
literature values. Our results for the growth constant $\mu$ and
the amplitudes $A$ and $D$ for the BCC and FCC lattices are the most
accurate ones to date.

The manuscript is organized as follows. First, in Sec.~\ref{sec:method} we
present a short outline of the length-doubling method, and present the
enumeration data. In Sec.~\ref{sec:analysis} we describe the analysis method we
use, before summarising our results and giving a brief conclusion in
Sec.~\ref{sec:conclusion}.

\section{Length-doubling method}
\label{sec:method}

We first present an intuitive description of the length-doubling method;
a more formal description can be found in~\cite{schram11}.
In the length-doubling method, the number $Z_{2N}$ of SAWs with a length
of $2N$ steps, with the middle rooted in the origin, is obtained from the
walks of length $N$, with one end rooted in the origin, and the number
$Z_N(S)$ of times that a subset $S$ of sites is visited by such a walk
of length $N$.  The lowest-order estimate for $Z_{2N}$ is the number of
combinations of two SAWs of length $N$, i.e. $Z_N^2$. This estimate is too
large since it includes pairs of SAWs which overlap. The first correction
to $Z_{2N}$ is the lowest-order estimate for the number of pairs of
overlapping SAWs, which can be obtained from the number $Z_N(\{s\})$
of SAWs of length $N$ which pass through a single site $s$. The first
correction is then to subtract $Z_N(\{s\})^2$, summed over all sites
$s$. This first correction is too large, as it includes pairs of SAWs
twice, if they intersect twice. The second correction corrects for
this over-subtraction, by adding the number $Z_N(\{s,t\})$ of SAWs that
pass through the pair of sites $\{s,t\}$.  Continuing this process with
groups of three sites, etc., the number $Z_{2N}$ of SAWs of length $2N$
can then be obtained by the length-doubling formula

\begin{equation}
\label{eqn:doubling}
Z_{2N} = Z_N^2 + \sum_{S \neq \emptyset} (-1)^{|S|} Z_N^2(S) ,
\end{equation}
where $|S|$ denotes the number of sites in $S$.

The usefulness of this formula lies in the fact that the numbers $Z_N(S)$ can
be obtained relatively efficiently:
\begin{itemize}
\item Generate each SAW of length $N$.
\item Generate for each SAW each of the $2^N$ subsets $S$ of lattice sites, and
increment the counter for each specific subset.
Multiple counters for the same subset $S$ must be avoided; this can
be achieved by sorting the sites within each subset in an unambiguous way.
\item Finally, compute the sum of the squares of these counters, with a positive
and negative sign for subsets with an even and odd number of sites, respectively,
as in Eq.~(\ref{eqn:doubling}).
\end{itemize}

As there are $Z_N$ walks of length $N$, each visiting $2^N$ subsets
of sites, the computational complexity is $\mathcal{O}(2^N Z_N) \sim
(2\mu)^N$ times some polynomial in $N$ which depends on implementation
details. This compares favourably to generating all $Z_{2N} \sim \mu^{2N}$
walks of length $2N$, provided $\mu > 2$. This is the case on the SC
lattice, with $\mu=4.684$, and even more so for the BCC and FCC lattices,
as we will show.  The length-doubling method can also compute the squared
end-to-end distance, summed over all SAW configurations; for details we
refer to~\cite{schram11}. Details on the efficient implementation
of this algorithm are presented in~\cite{schram13}.

The direct results of the length-doubling method, applied to SAWs on the BCC and
FCC lattices, are presented in Tables 1 and 2, respectively.
The BCC results for $N \leq 26$ and FCC results for $N \leq 22$ were obtained
and verified by two independent computer programs:
SAWdoubler 2.0, available from \url{http://www.staff.science.uu.nl/~bisse101/SAW/},
and Raoul Schram's program.
The BCC results presented  for the largest problems $N=27,28$ were obtained by SAWdoubler 2.0 only, and
the FCC results for $N=23,24$ were obtained by Raoul Schram's program only.
Thus the largest two problem instances for each lattice were not independently verified since
these require a very large amount of computer time and memory.
Still, based on our analysis we believe that the given values are correct. 
\begin{table}
\caption{
\label{tab:bcc}
Enumeration results for the number of three-dimensional
self-avoiding walks $Z_N$ and the sum of their squared end-to-end distances $P_N$
on the BCC lattice.}
\begin{tabular}{rrr}
\hline
$N$ & $Z_N$ & $P_N$ \\
\hline
1&8\,&\,24\,\\
2&56\,&\,384\,\\
3&392\,&\,4\,248\,\\
4&2\,648\,&\,40\,704\,\\
5&17\,960\,&\,358\,008\,\\
6&120\,056\,&\,2\,987\,232\,\\
7&804\,824\,&\,23\,999\,880\,\\
8&5\,351\,720\,&\,187\,661\,376\,\\
9&35\,652\,680\,&\,1\,436\,494\,872\,\\
10&236\,291\,096\,&\,10\,816\,140\,768\,\\
11&1\,568\,049\,560\,&\,80\,339\,567\,112\,\\
12&10\,368\,669\,992\,&\,590\,168\,152\,512\,\\
13&68\,626\,647\,608\,&\,4\,294\,543\,350\,696\,\\
14&453\,032\,542\,040\,&\,31\,003\,097\,851\,872\,\\
15&2\,992\,783\,648\,424\,&\,222\,268\,142\,153\,784\,\\
16&19\,731\,335\,857\,592\,&\,1\,583\,984\,756\,900\,544\,\\
17&130\,161\,040\,083\,608\,&\,11\,228\,345\,566\,400\,136\,\\
18&857\,282\,278\,813\,256\,&\,79\,223\,666\,339\,548\,320\,\\
19&5\,648\,892\,048\,530\,888\,&\,556\,634\,161\,952\,309\,400\,\\
20&37\,175\,039\,569\,217\,672\,&\,3\,896\,382\,415\,388\,139\,840\,\\
21&244\,738\,250\,638\,121\,768\,&\,27\,181\,650\,674\,871\,447\,672\,\\
22&1\,609\,522\,963\,822\,562\,936\,&\,189\,042\,890\,267\,974\,827\,744\,\\
23&10\,588\,362\,063\,533\,857\,304\,&\,1\,311\,064\,323\,033\,684\,408\,072\,\\
24&69\,595\,035\,470\,413\,829\,144\,&\,9\,069\,398\,712\,299\,296\,227\,648\,\\
25&457\,555\,628\,726\,692\,288\,712\,&\,62\,590\,336\,418\,536\,387\,660\,248\,\\
26&3\,005\,966\,051\,800\,541\,943\,464\,&\,431\,019\,462\,253\,450\,273\,360\,416\,\\
27&19\,752\,610\,526\,081\,274\,414\,584&\,2\,962\,188\,249\,772\,759\,155\,770\,280\,\\
28&129\,713\,248\,317\,927\,812\,262\,200&\,20\,319\,964\,852\,485\,237\,389\,626\,176\,\\
\hline
\end{tabular}
\end{table}

\begin{table}
\caption{
\label{tab:fcc}
Enumeration results for the number of three-dimensional
self-avoiding walks $Z_N$ and the sum of their squared end-to-end distances $P_N$
on the FCC lattice.}
\begin{tabular}{rrr}
\hline
$N$ & $Z_N$ & $P_N$ \\
\hline
1&12\,&\,24\,\\
2&132\,&\,576\,\\
3&1\,404\,&\,9\,816\,\\
4&14\,700\,&\,144\,288\,\\
5&152\,532\,&\,1\,951\,560\,\\
6&1\,573\,716\,&\,25\,021\,536\,\\
7&16\,172\,148\,&\,309\,080\,808\,\\
8&165\,697\,044\,&\,3\,714\,659\,040\,\\
9&1\,693\,773\,924\,&\,43\,714\,781\,448\,\\
10&17\,281\,929\,564\,&\,505\,948\,384\,608\,\\
11&176\,064\,704\,412\,&\,5\,777\,220\,825\,912\,\\
12&1\,791\,455\,071\,068\,&\,65\,234\,797\,723\,584\,\\
13&18\,208\,650\,297\,396\,&\,729\,724\,191\,726\,408\,\\
14&184\,907\,370\,618\,612\,&\,8\,097\,639\,351\,530\,304\,\\
15&1\,876\,240\,018\,679\,868\,&\,89\,239\,258\,469\,121\,912\,\\
16&19\,024\,942\,249\,966\,812\,&\,977\,545\,487\,795\,069\,952\,\\
17&192\,794\,447\,005\,403\,916\,&\,10\,651\,662\,728\,070\,257\,016\,\\
18&1\,952\,681\,556\,794\,601\,732\,&\,115\,520\,552\,778\,504\,791\,136\,\\
19&19\,767\,824\,914\,170\,222\,996\,&\,1\,247\,619\,751\,507\,795\,906\,248\,\\
20&200\,031\,316\,330\,580\,106\,948\,&\,13\,423\,705\,093\,594\,869\,393\,216\,\\
21&2\,023\,330\,401\,919\,804\,218\,996\,&\,143\,942\,374\,595\,787\,212\,970\,696\,\\
22&20\,458\,835\,772\,261\,851\,432\,748\,&\,1\,538\,749\,219\,442\,520\,114\,999\,744\,\\
23&206\,801\,586\,042\,610\,941\,719\,148\,&\,16\,403\,200\,314\,230\,418\,676\,555\,512\,\\
24&2\,089\,765\,228\,215\,904\,826\,153\,292\,&\,174\,411\,223\,302\,510\,038\,302\,309\,440\,\\
\hline
\end{tabular}
\end{table}

\section{Analysis}
\label{sec:analysis}
We now proceed to analyse our series in order to extract estimates for
various parameters.
In addition to the expressions for $Z_N$ and $P_N / Z_N$ in
Eqs~(\ref{eq:zn}) and (\ref{eq:rhon}), we also have
\begin{align}
    P_N &= \sigma A D \mu^N N^{2\nu + \gamma-1} \left(1 + 
    \frac{c}{N^{\Delta_1}} + O\left(\frac{1}{N}\right)\right). \label{eq:pn}
\end{align}
As discussed earlier, we expect the critical exponents 
$\gamma$ and $\nu$ and the leading correction-to-scaling exponent $\Delta_1$
to
be the same for self-avoiding walks on the SC, BCC, and FCC
lattices. 
The amplitudes $A$ and $D$ are non-universal quantities, i.e. they are
lattice dependent, while $\sigma = 2$ for the BCC lattice and $\sigma =
3$ for the FCC lattice. In the analysis below, we include a subscript to
indicate the appropriate lattice.

The BCC lattice is bipartite, which introduces an additional 
competing
correction which has a factor of $(-1)^N$, so causing odd-even
oscillations. 
We reduce the influence of this additional sub-leading
correction by separately treating the sequences for even and odd $N$.
See \cite{clisby07} for more detailed discussion
on this point for the asymptotic behaviour of $Z_N$ on the 
SC lattice, which is also bipartite.

We now describe the method of analysis we used, which involved two
stages: extrapolation of the series via a recently introduced method involving
differential approximants~\cite{guttmann16}, and then direct fitting of
the extended series with the asymptotic forms in Eqs~(\ref{eq:zn}), (\ref{eq:rhon}), and (\ref{eq:pn}). 
We report our final estimates
in Table~\ref{tab:summary} at the end of the section.

\subsection{Extrapolation}

Perhaps the most powerful general-purpose method for the analysis of
series arising from lattice models in statistical mechanics is the
method of differential approximants, described in \cite{guttmann89b}.
The basic idea is to approximate
the unknown generating function $F$ by the solution of an
ordinary differential equation with polynomial coefficients.
In
particular if we know $r$ coefficients $f_0,
f_1, \cdots, f_{r-1}$ of our generating function $F$,
then we can determine polynomials $Q_i(z)$ and
$P(z)$ which satisfy the following $K$th order differential equation order by order:
\begin{align}
    \sum_{i=0}^{K} Q_i(z) \left(z \frac{d}{dz}\right)^i F(z) = P(z).
\end{align}
The function determined by the resulting differential equation is our
approximant. 
The power of the method derives from the fact that such ordinary
differential equations accommodate the kinds of critical behaviour that
are typically seen for models of interest.

Differential approximants are extremely effective at extracting
information about critical exponents from the long series that have been
obtained for two-dimensional lattice models, such as self-avoiding
polygons~\cite{Clisby2012newtransfermatrix} or walks on the square
lattice~\cite{Jensen2016SquareLatticeSAWsBiasedDifferentialApproximants}.
However, differential approximants have been far less successful for
the shorter series available for three-dimensional models such as 
SAWs on the simple cubic
lattice~\cite{clisby07,schram11}. For short series, it seems that
corrections-to-scaling due to confluent corrections are too strong at
the orders that can be reached to be able to reliably determine critical
exponents. (In fact, it is extremely easy to be misled by apparent
convergence, while in fact estimates have not settled down to their
asymptotic values.) The method that has proved most reliable is direct
fitting of the asymptotic form~\cite{clisby07}, which we describe in
the next sub-section.

However, we can do better than the usual method of performing direct
fits of the original series, and
adopt a promising new approach recently invented by Guttmann~\cite{guttmann16}, 
which is a hybrid of the differential
approximant and direct fitting techniques.
The underlying idea is to exploit the fact that differential approximants can
be used to extrapolate series with high accuracy even in circumstances
when the resulting estimates for critical exponents are not particularly
accurate, or even when the asymptotic behaviour is non-standard such as
being of stretched exponential form.
The extrapolations can be extremely useful in cases where
corrections-to-scaling are large, as the few extra terms they provide
may be the only evidence of a clear trend from the direct fits.

We have 28 exact terms for the BCC series, and 24 exact terms for the
FCC series.  We used second order inhomogeneous approximants to
extrapolate the series for $Z_N$, $P_N$, and $P_N/Z_N$, where we allowed
the multiplying polynomials to differ by degree at most 3. In each case
we calculated trimmed mean values, eliminating the outlying top and
bottom 10\% of estimates, with the standard deviation of the remaining
extrapolated coefficients providing a proxy for the confidence interval.
Note that this is an assumption, and relies on the extrapolation
procedure working well for our problem. In practice, this approach of
inferring the confidence interval from the spread of estimates appears
to be quite reliable in the cases for which it has been tested.  We have
also confirmed the reliability of the extrapolations by using the method
to ``predict'' known coefficients from truncated series.  We report our
extended series in Tables~\ref{tab:bccextrapolation} and
\ref{tab:fccextrapolation}.

\begin{table}
\caption{
\label{tab:bccextrapolation}
Extrapolated coefficients of the various BCC series obtained from differential
    approximants. The confidence intervals are the standard deviations of
    the central 80\% of estimates.}
\begin{tabular}{llll}
\hline
    $N$ & $Z_N$ & $P_N$ & $P_N/Z_N$ \\
    \hline
29 & 8.51984378150(70)\e{23} & 1.39148952051(11)\e{26} & 163.323360851(42) \\
30 & 5.5928669767(12)\e{24} & 9.5134610227(17)\e{26} & 170.09989796(10) \\
31 & 3.6720987764(23)\e{25} & 6.4944301898(72)\e{27} & 176.85880953(40) \\
32 & 2.4097907972(39)\e{26} & 4.4272318727(75)\e{28} & 183.71851486(77) \\
33 & 1.5816583535(44)\e{27} & 3.014025691(25)\e{29} & 190.5611070(19) \\
34 & 1.037661297(10)\e{28} & 2.049378203(42)\e{30} & 197.4997221(33) \\
35 & 6.808628821(74)\e{28} & 1.391831542(69)\e{31} & 204.4217013(47) \\
36 & 4.46574383(26)\e{29} & 9.44216466(95)\e{31} & 211.435420(11) \\
37 & 2.929428561(97)\e{30} & 6.3988380(13)\e{32} & 218.432947(13) \\
38 & 1.9209657(36)\e{31} & 4.3321295(17)\e{33} & 225.518346(32) \\
\hline
\end{tabular}
\end{table}

\begin{table}
\caption{
\label{tab:fccextrapolation}
Extrapolated coefficients of the various FCC series obtained from differential
    approximants. The confidence intervals are the standard deviations of
    the central 80\% of estimates.}
\begin{tabular}{llll}
\hline
    $N$ & $Z_N$ & $P_N$ & $P_N/Z_N$ \\
    \hline
25 & 2.1111652709103(46)\e{25} & 1.85010449211473(82)\e{27} & 87.63428034806(26) \\
26 & 2.132245848773(38)\e{26}  & 1.9582778101818(72)\e{28}  & 91.8410891195(22) \\
27 & 2.15303362972(17)\e{27}   & 2.068615279889(35)\e{29}   & 96.079097491(11) \\
28 & 2.1735525326(10)\e{28}    & 2.18110187619(13)\e{30}    & 100.347327420(41) \\
29 & 2.1938240975(32)\e{29}    & 2.29572427539(38)\e{31}    & 104.64486552(13) \\
30 & 2.2138677922(93)\e{30}    & 2.41247069749(92)\e{32}    & 108.97085672(37) \\
31 & 2.233701285(63)\e{31}     & 2.5313307684(21)\e{33}     & 113.32449876(98) \\
32 & 2.25334058(14)\e{32}      & 2.6522953987(45)\e{34}     & 117.7050374(24) \\
33 & 2.2728013(51)\e{33}       & 2.7753566769(86)\e{35}     & 122.1117622(56) \\
\hline
\end{tabular}
\end{table}

\subsection{Direct fits}

We then fitted sequences of consecutive terms of the extrapolated series
for $Z_N$ and $P_N/Z_N$ to the asymptotic forms given in Eqs~(\ref{eq:zn})
and (\ref{eq:rhon}), respectively. We found that fits of $P_N/Z_N$ were
superior to fits of $P_N$ for estimates of $\nu$ and the parameter $D$,
and hence we do not report fits of $P_N$ here.

To convert the fitting problem to a linear equation, we took the
logarithm of the coefficients, which from
Eqs~(\ref{eq:zn}) and (\ref{eq:rhon}) we expect to have the following
asymptotic forms:
\begin{align}
    \log Z_N &= N \log \mu  +  (\gamma-1) \log N + \log A + \frac{a}{N^{\Delta_1}} +
    O\left(\frac{1}{N}\right); \label{eq:logzn} \\
    \log \frac{P_N}{Z_N} &= {2\nu} \log N + \log \sigma D +  \frac{b}{N^{\Delta_1}} +
    O\left(\frac{1}{N}\right) \label{eq:logrhon}
\end{align}
We used the linear fitting routine ``lm'' in the statistical
programming language R to perform the fits.

In all of the fits, we biased the exponent $\Delta_1$ of the leading
correction-to-scaling term, performing the fits for three different
choices of $\Delta_1 = 0.520, 0.528, 0.536$ which correspond to the best
Monte Carlo estimate of $\Delta_1 = 0.528(8)$.
We approximated the next-to-leading correction-to-scaling term with a
term of order $1/N$, which we expect to behave as an effective term
which takes into account three competing corrections with exponents
$-2\Delta_1, -1, -\Delta_2 \approx -1$.
For $\log Z_N$, we fitted $\log A$,
$\log \mu$, $\gamma$, the amplitude $a$, and the
amplitude of the $1/N$ effective term.
For $\log (P_N/Z_N)$, we fitted $ \log D$,
$\nu$, the amplitude $b$, 
and the 
amplitude of the $1/N$ term.
For the BCC lattice, we minimised the impact of the odd-even oscillations
by fitting even and odd subsequences separately.
We included the extrapolated coefficients in our fits, repeating the
calculation for the central estimates and for values which are one standard
deviation above and below them.

This procedure gave us up to nine estimates for each sequence of
coefficients (from the three choices of $\Delta_1$, and the three
choices of extrapolated coefficient values). For the central parameter
estimates we used the case where $\Delta_1 = 0.528$ (the central value)
in combination with the central
value of the extrapolated coefficients. We also calculated the maximum
and minimum parameter estimates over the remaining 8 cases.

For the BCC lattice, we found that 5 of the extrapolated coefficients
gave a spread which was only moderately greater than the spread arising
from varying $\Delta_1$, effectively extending the series to 33 terms.
For the FCC lattice, we found we could use 3
additional coefficients, extending the series to 27 terms.

For each of the parameter estimates, we plotted them against the
expected relative magnitude of the first neglected correction-to-scaling
term. This should result in approximately linear convergence as we
approach the $N \rightarrow \infty$ limit which corresponds to
approaching the $y$-axis from the right in the following figures. In
Eqs~(\ref{eq:logzn}) and (\ref{eq:logrhon}) we expect that the next
term, which is not included in the fits, is of $O(N^{-1-\Delta_1})$;
given that $\Delta_1 \approx 0.5$, we take the neglected term to be
$O(N^{-3/2})$. The value of $N$ that is used in the plot is the maximum
value of $N$ in the sequence of fitted coefficients, which we denote
$N_{\max}$ in the plots.

\begin{figure}[htb]
\begin{center}
\begin{minipage}{0.45\textwidth}
\begin{center}
    \includegraphics[width=1.0\textwidth]{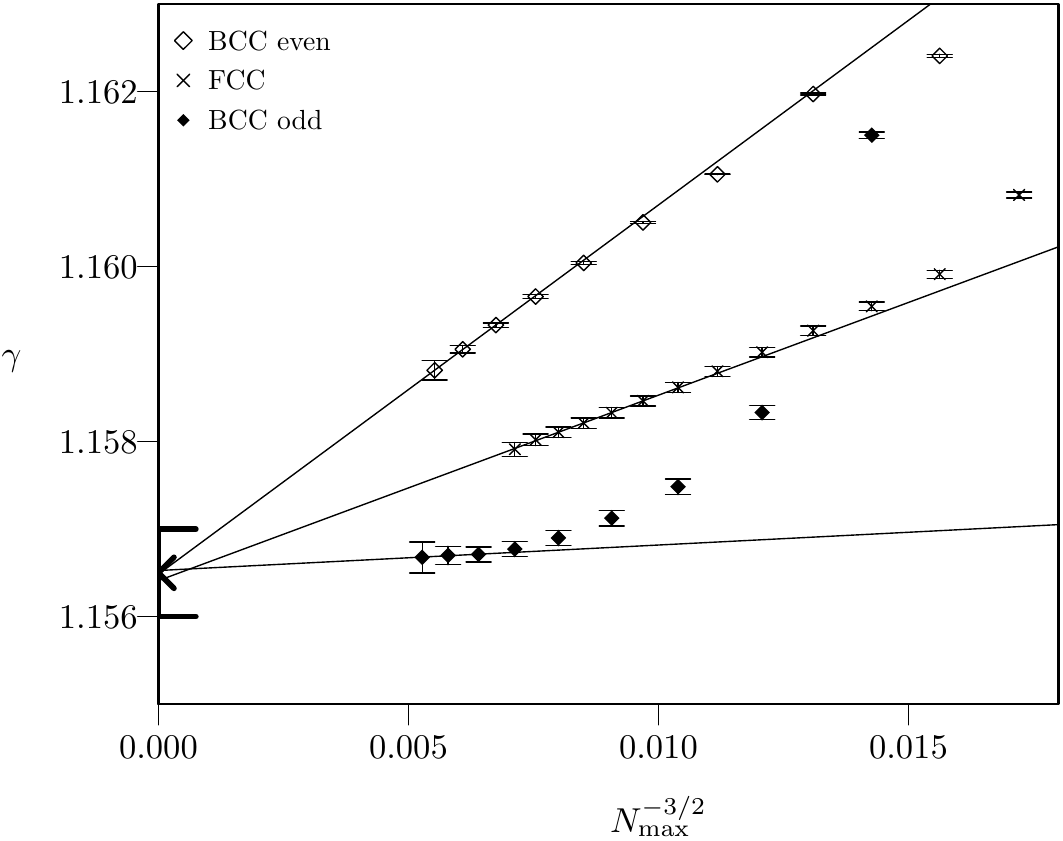}
\end{center}
\vspace{-4ex}
  \caption{Variation of fitted value of $\gamma$ with
    $N_{\max}$.
    The line of best fit to the final six values is shown for the FCC
    lattice and to the final three
    values for the BCC lattice, separately for the odd and even
    values. Our final estimate is plotted on the $y$-axis.}
  \label{fig:gamma}
\end{minipage}
\hspace{2em}
\begin{minipage}{0.45\textwidth}
\begin{center}
    \includegraphics[width=1.0\textwidth]{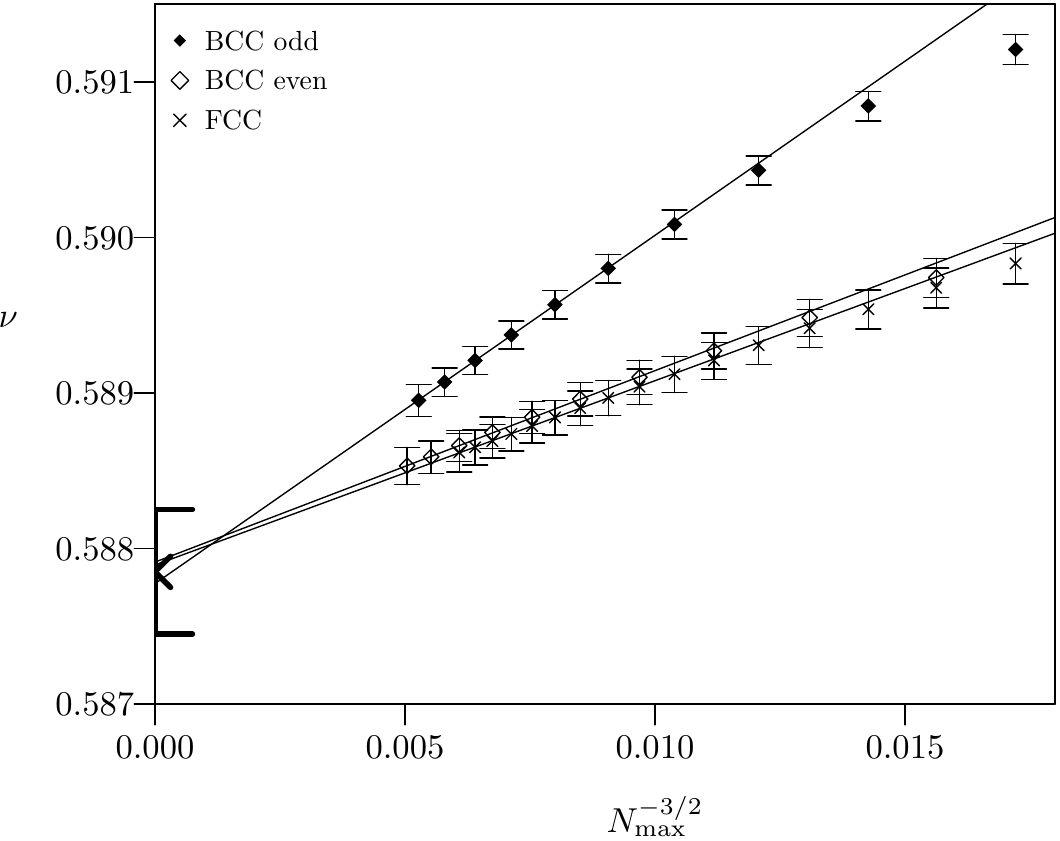}
\end{center}
\vspace{-4ex}
  \caption{Variation of fitted value of $\nu$ with
    $N_{\max}$.
    The line of best fit to the final six values is shown for the FCC
    lattice and to the final
    three  values for the BCC lattice, separately for the odd and even
    values. Our final estimate is plotted on the $y$-axis.}
  \label{fig:nu}
\end{minipage}
\end{center}
\end{figure}

\begin{figure}[htb]
\begin{center}
\begin{minipage}{0.45\textwidth}
\begin{center}
    \includegraphics[width=1.0\textwidth]{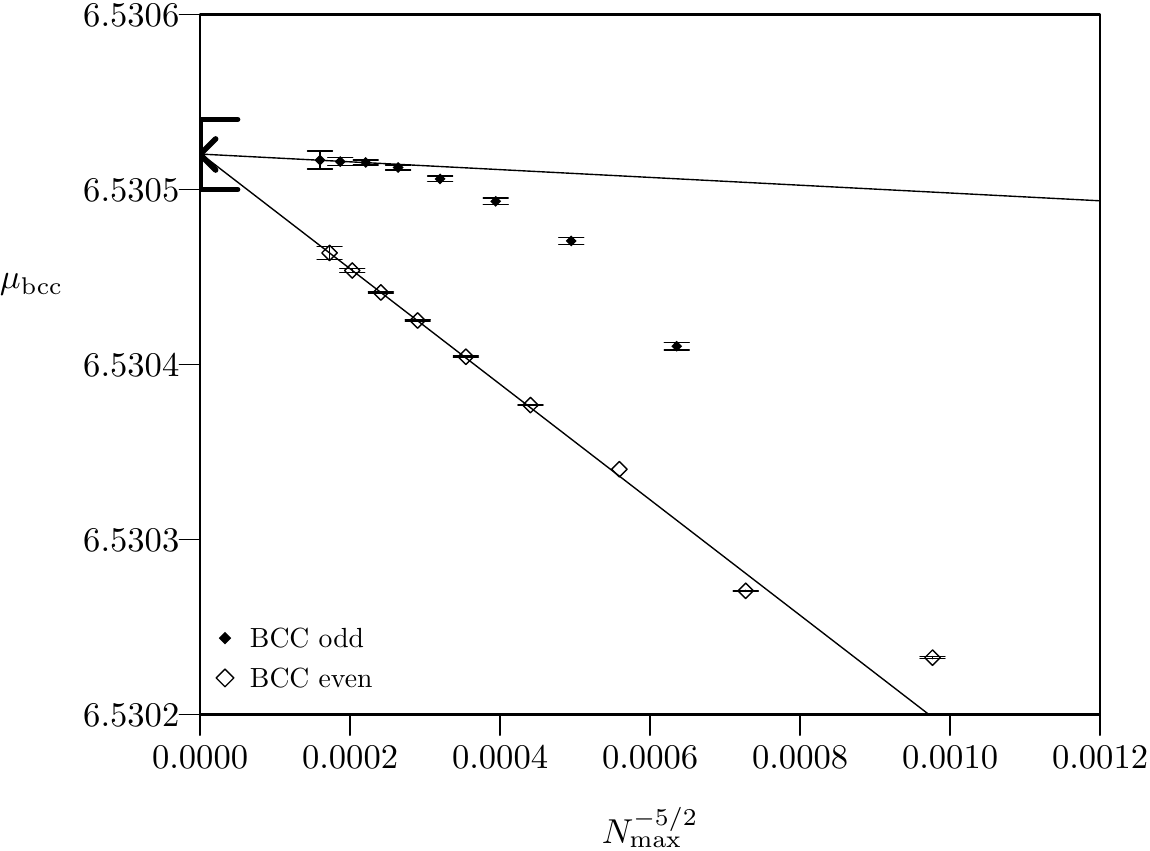}
\end{center}
\vspace{-4ex}
    \caption{Variation of fitted value of $\mu_{\rm bcc}$ with
    $N_{\max}$.
    The line of best fit 
    3 values is shown, separately for the odd and even
    values.
    Our final estimate is plotted on the $y$-axis.}
  \label{fig:mubcc}
\end{minipage}
\hspace{2em}
\begin{minipage}{0.45\textwidth}
\begin{center}
    \includegraphics[width=1.0\textwidth]{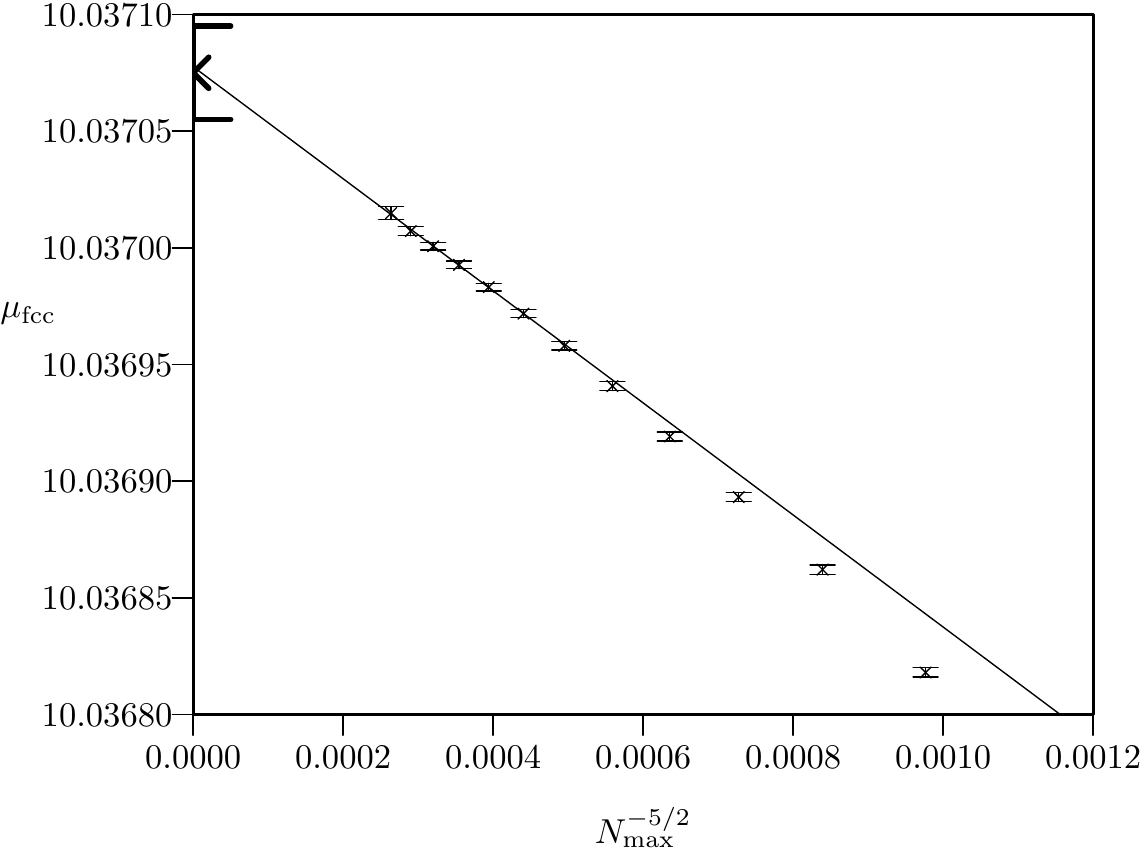}
\end{center}
\vspace{-4ex}
    \caption{Variation of fitted value of $\mu_{\rm fcc}$ with
    $N_{\max}$.
    The line of best fit to the final six values is shown.
    Our final estimate is plotted on the $y$-axis.}
  \label{fig:mufcc}
\end{minipage}
\end{center}
\end{figure}

\begin{figure}[htb]
\begin{center}
\begin{minipage}{0.45\textwidth}
\begin{center}
    \includegraphics[width=1.0\textwidth]{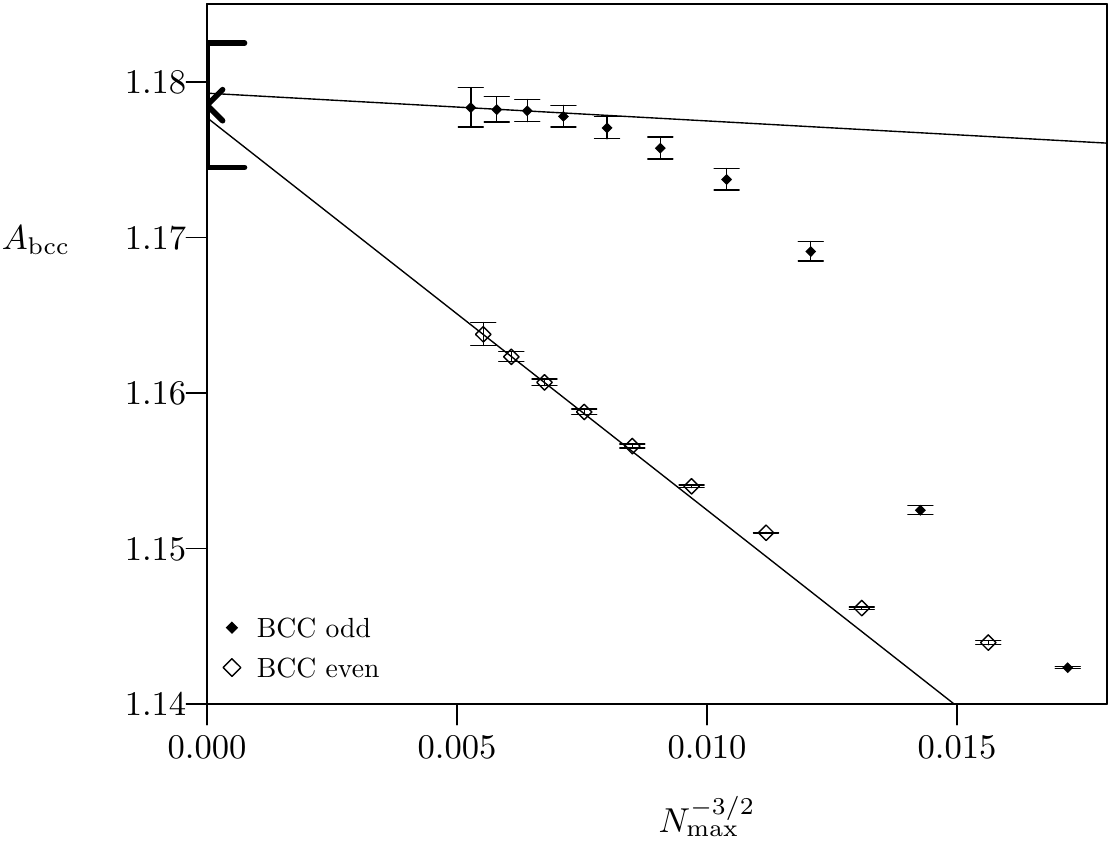}
\end{center}
\vspace{-4ex}
    \caption{Variation of fitted value of $A_{\rm bcc}$ with
    $N_{\max}$.
    The line of best fit 
    3 values is shown, separately for the odd and even
    values.
    Our final estimate is plotted on the $y$-axis.}
  \label{fig:Abcc}
\end{minipage}
\hspace{2em}
\begin{minipage}{0.45\textwidth}
\begin{center}
    \includegraphics[width=1.0\textwidth]{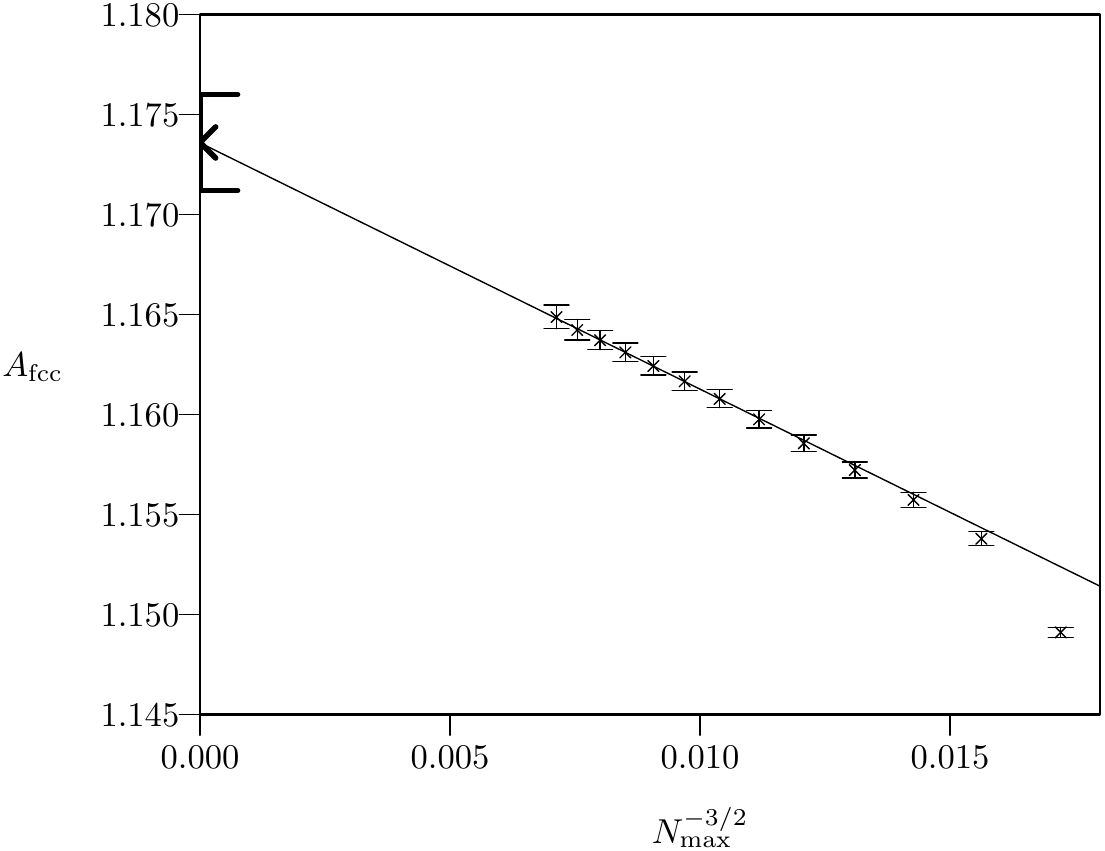}
\end{center}
\vspace{-4ex}
    \caption{Variation of fitted value of $A_{\rm fcc}$ with
    $N_{\max}$.
    The line of best fit to the final six values is shown.
    Our final estimate is plotted on the $y$-axis.}
  \label{fig:Afcc}
\end{minipage}
\end{center}
\end{figure}

\begin{figure}[htb]
\begin{center}
\begin{minipage}{0.45\textwidth}
\begin{center}
    \includegraphics[width=1.0\textwidth]{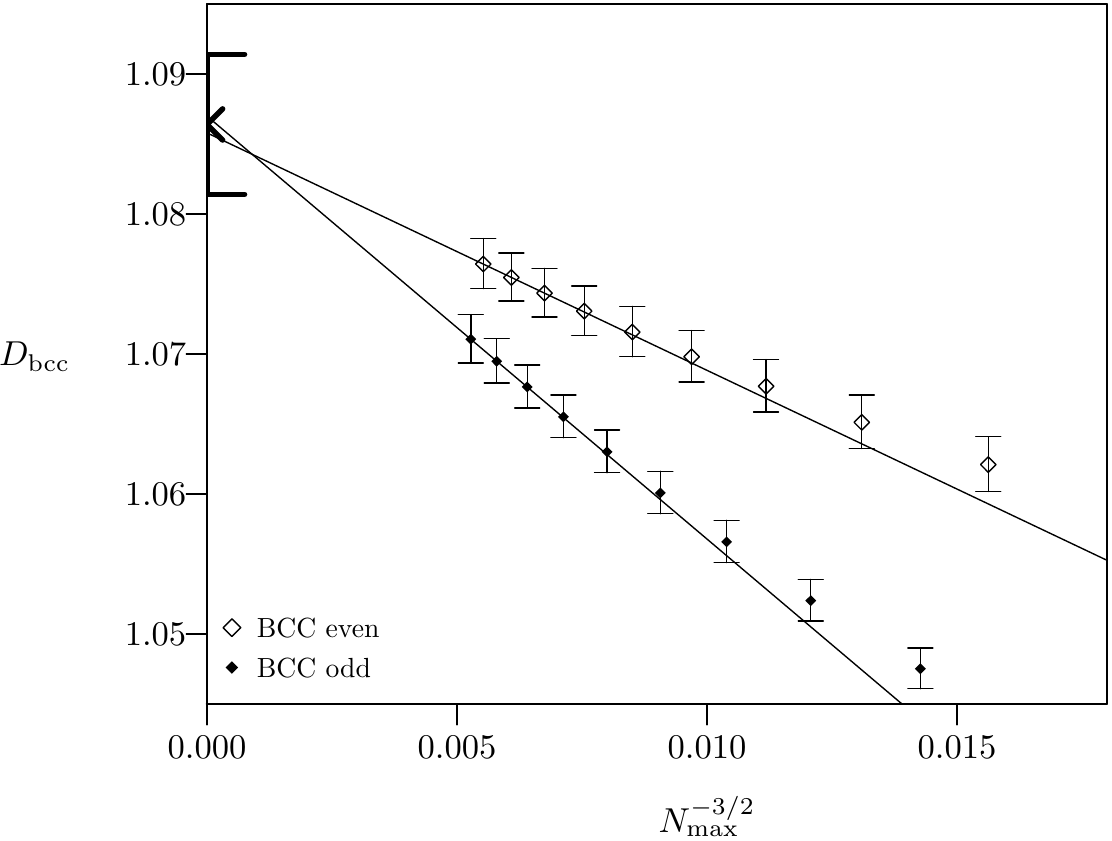}
\end{center}
\vspace{-4ex}
    \caption{Variation of fitted value of $D_{\rm bcc}$ with
    $N_{\max}$.
    The line of best fit 
    3 values is shown, separately for the odd and even
    values.
    Our final estimate is plotted on the $y$-axis.}
  \label{fig:Dbcc}
\end{minipage}
\hspace{2em}
\begin{minipage}{0.45\textwidth}
\begin{center}
    \includegraphics[width=1.0\textwidth]{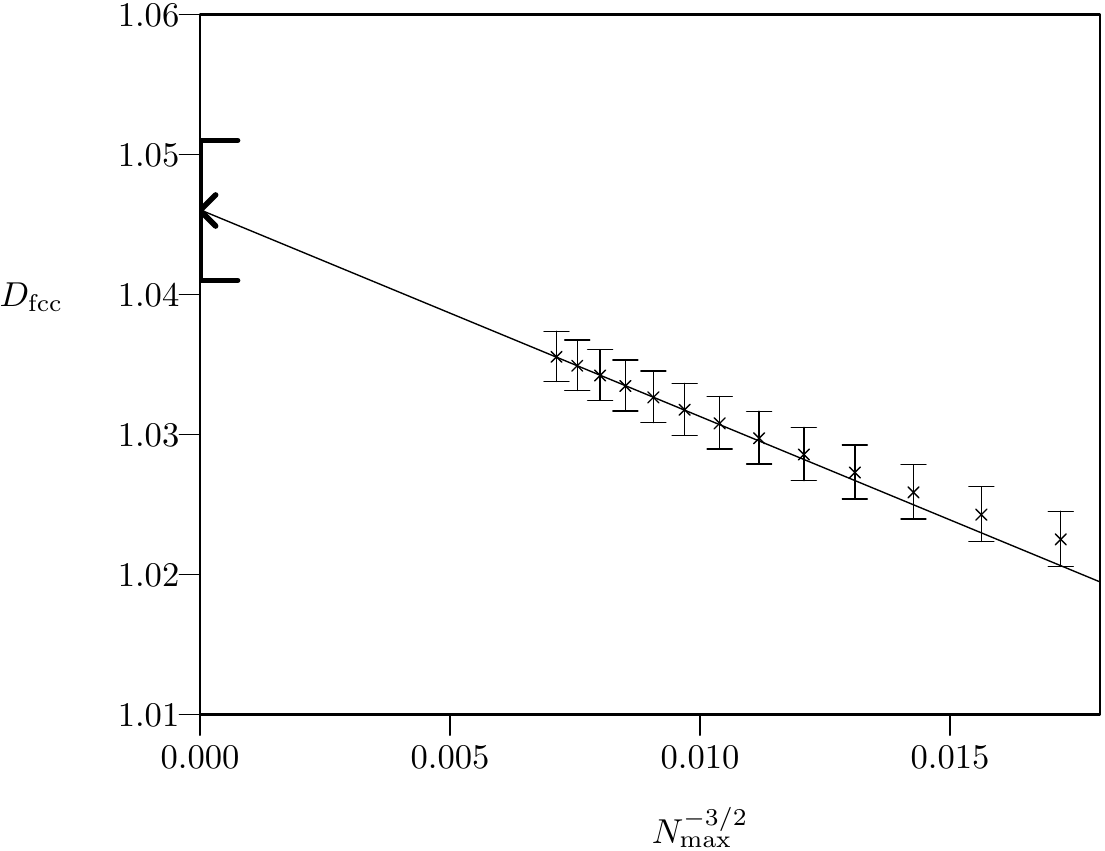}
\end{center}
\vspace{-4ex}
    \caption{Variation of fitted value of $D_{\rm fcc}$ with
    $N_{\max}$.
    The line of best fit to the final six values is shown.
    Our final estimate is plotted on the $y$-axis.}
  \label{fig:Dfcc}
\end{minipage}
\end{center}
\end{figure}

We plot our fitted values in Figures~\ref{fig:gamma}--\ref{fig:Dfcc}. 
For ease of interpretation we converted estimates of $\log \mu$, $\log
A$, and $\log D$ to estimates of $\mu$, $A$, and $D$.
We
note that the parameter estimates arising from the odd subsequence of
the BCC series for $Z_N$ benefited dramatically from the extrapolated
sequence. Examining estimates for $\gamma$ in Fig.~\ref{fig:gamma},
$\mubcc$ in Fig.~\ref{fig:mubcc}, and $\Abcc$ in Fig.~\ref{fig:Abcc} we
see in each case that the trend of the odd subsequence would be
dramatically different were it not for the three additional odd terms in the
extrapolated sequence. In other cases the additional coefficients are
useful, and certainly make the trend for the estimates clearer, but are
not as crucial.

Our final parameter estimates are plotted on the $y$-axes.

\section{Summary and conclusion}
\label{sec:conclusion}

We give our estimates for $\gamma$ and $\nu$ in Table~\ref{tab:summary}, where
we also include estimates coming from the literature.  We observe that our
estimates are consistent with the literature values, but that the
recent Monte Carlo estimates of $\gamma$ and $\nu$, using the pivot algorithm,
are far more accurate than the estimates from series.  The estimates coming from
our enumerations on the BCC and FCC lattices are not quite as precise as the
estimates coming from the SC lattice only, but the fact that they are coming
from two independent sources, with different systematic errors, makes these new
estimates more robust.

In addition, our estimates of the non-universal quantities for the BCC lattice
are $\Abcc = \ourAbcc$, $\Dbcc = \ourDbcc$, 
and $\mubcc = \ourmubcc$, which should be compared with earlier estimates of 6.5304(13)~\cite{guttmann89} from 1989, and
unbiased and biased estimates respectively of 6.53036(9) and
6.53048(12)~\cite{Butera1997Nvectorspin} from 1997.
Our estimates of the non-universal quantities for the FCC lattice
are $\Afcc = \ourAfcc$, $\Dfcc = \ourDfcc$, and $\mufcc = \ourmufcc$, which should be compared with earlier estimates of 10.03655~\cite{McKenzie1979SAWsFCC} from 1979, and
10.0364(6)~\cite{guttmann87} from 1987 (where these
estimates come from different analyses of the same $N \leq 14$ term
series).

\begin{table}
\caption{
\label{tab:summary}
    Summary of our parameter estimates for $\gamma$ and $\nu$, with comparison to values from the
    literature. Except where noted, the series estimates for $\gamma$ and $\nu$ from the
    literature come from the simple cubic lattice.}
\begin{tabular}{lll}
    \hline
Source\footnote{Abbreviations: MC $\equiv$ Monte
    Carlo, CB $\equiv$ conformal bootstrap, FT $\equiv$ field theory, 
    MCRG $\equiv$ Monte Carlo renormalization group.} 
    &  $\gamma$ & $\nu$\\
    \hline
    This work &  $\ourgamma$ &  $\ournu$\\
    \cite{Clisby2017Scale-freeGammaSAWsArxiv} MC (2017) & 1.15695300(95) & \\
    \cite{Clisby2016HydrodynamicRadiusForSAWs} MC (2016) &  & 0.58759700(40) \\
    \cite{Shimada2016SAWsIn3DConformalBootstrap} CB (2016) & 1.1588(25) & 0.5877(12) \\
     \cite{schram11} Series $N \leq 36$ (2011) & 1.15698(34) & 0.58772(17)\\
    \cite{Clisby2010AccurateEstimateCritical} MC (2010) & & 0.587597(7) \\
     \cite{clisby07}\footnote{Using Eqs (74) and (75) with $0.516 \leq \Delta_1 \leq 0.54$.} Series $N \leq 30$ (2007) & 1.1569(6) & 0.58774(22)\\
     \cite{Hsu2004Polymersconfinedbetween} MC (2004)&1.1573(2) & \\
     \cite{Prellberg2001Scalingselfavoiding} MC (2001) & & 0.5874(2) \\
     \cite{macdonald00} Series $N \leq 26$ (2000) & 1.1585 & 0.5875\\
     \cite{Caracciolo1998Highprecisiondetermination} MC (1998) & 1.1575(6) \\
     \cite{Guida1998CriticalexponentsN} FT $d=3$ (1998) & 1.1596(20) & 0.5882(11)\\
     \cite{Guida1998CriticalexponentsN} FT $\epsilon$ bc (1998) & 1.1571(30) & 0.5878(11) \\
     \cite{Butera1997Nvectorspin} Series $N \leq 21$ (1997) & 1.161(2) & 0.592(2) \\
     \cite{Butera1997Nvectorspin} Series $N \leq 21$, biased (1997) & 1.1594(8) & 0.5878(6) \\
     \cite{Butera1997Nvectorspin} BCC series $N \leq 21$ (1997) & 1.1612(8) & 0.591(2) \\
     \cite{Butera1997Nvectorspin} BCC series $N \leq 21$, biased (1997) & 1.1582(8) & 0.5879(6) \\
     \cite{Belohorec1997Renormalizationgroupcalculation} MCRG (1997) & & 0.58756(5) \\
     \cite{Li1995CriticalExponentsHyperscaling} MC (1995)  & & 0.5877(6) \\
     \cite{macdonald92} Series $N \leq 23$ (1992)& 1.16193(10) & \\
     \cite{guttmann89} Series $N \leq 21$ (1989) & 1.161(2) & 0.592(3) \\
    \hline
\end{tabular}
\end{table}

In conclusion, the length-doubling algorithm has resulted in significant
extensions of the BCC and FCC series. The application of a recently invented
series analysis technique~\cite{guttmann16}, which combines series
extrapolation from differential approximants with direct fitting of the
extrapolated series,
has given excellent estimates of the various critical parameters.
In particular, estimates of the growth constants for the BCC and FCC
lattices are far more accurate than the previous literature values.

\section*{Acknowledgements}

This work was sponsored by NWO-Science for the use of supercomputer
facilities under the project SH-349-15.  Computations were carried out
on the Cartesius supercomputer at SURFsara in Amsterdam.  N.C.
acknowledges support from the Australian Research Council under the
Future Fellowship scheme (project number FT130100972) and Discovery
scheme (project number DP140101110).


\begin{thebibliography}{32}
\providecommand{\natexlab}[1]{#1}
\providecommand{\url}[1]{\texttt{#1}}
\expandafter\ifx\csname urlstyle\endcsname\relax
  \providecommand{\doi}[1]{doi: #1}\else
  \providecommand{\doi}{doi: \begingroup \urlstyle{rm}\Url}\fi

\bibitem[Madras and Slade(1993)]{madras93}
Neal Madras and Gordon Slade.
\newblock \emph{The Self-Avoiding Walk}.
\newblock Probability and its applications. {Birkh\"{a}user}, Boston, MA, 1993.

\bibitem[Janse~van Rensburg(2015)]{janse15}
E.~J. Janse~van Rensburg.
\newblock \emph{The Statistical Mechanics of Interacting Walks, Polygons,
  Animals and Vesicles}.
\newblock Oxford University Press, Oxford, UK, second edition, 2015.

\bibitem[Orr(1947)]{orr47}
W.~J.~C Orr.
\newblock Statistical treatment of polymer solutions at infinite dilution.
\newblock \emph{Transactions Faraday Society}, 43:\penalty0 12--27, 1947.

\bibitem[Fisher and Sykes(1959)]{fisher59}
Michael~E. Fisher and M.~F. Sykes.
\newblock Excluded-volume problem and the {Ising} model of ferromagnetism.
\newblock \emph{Physical Review}, 114:\penalty0 45--58, 1959.

\bibitem[Sykes(1961)]{Sykes1961SomeCountingTheorems}
M.~F. Sykes.
\newblock Some counting theorems in the theory of the {I}sing model and the
  excluded volume problem.
\newblock \emph{J. Math. Phys.}, 2:\penalty0 52--62, 1961.

\bibitem[Sykes(1963)]{Sykes1963SelfAvoidingWalksontheSimpleCubicLattice}
M.~F. Sykes.
\newblock Self avoiding walks on the simple cubic lattice.
\newblock \emph{J. Chem. Phys.}, 39:\penalty0 410--412, 1963.

\bibitem[Sykes et~al.(1972)Sykes, Guttmann, Watts, and
  Roberts]{Sykes1972SAWsAndSARsOnVariousLattices}
M.~F. Sykes, A.~J. Guttmann, M.~G. Watts, and P.~D. Roberts.
\newblock The asymptotic behaviour of selfavoiding walks and returns on a
  lattice.
\newblock \emph{J. Phys. A: Gen. Phys.}, 5:\penalty0 653--660, 1972.

\bibitem[Guttmann(1987)]{guttmann87}
A.~J. Guttmann.
\newblock On the critical behaviour of self-avoiding walks.
\newblock \emph{J. Phys. A: Math. Gen.}, 20:\penalty0 1839--1854, 1987.

\bibitem[Guttmann(1989{\natexlab{a}})]{guttmann89}
A.~J. Guttmann.
\newblock On the critical behaviour of self-avoiding walks: {$II$}.
\newblock \emph{J. Phys. A: Math. Gen.}, 22:\penalty0 2807--2813,
  1989{\natexlab{a}}.

\bibitem[MacDonald et~al.(1992)MacDonald, Hunter, Kelly, and Jan]{macdonald92}
D.~MacDonald, D.~L. Hunter, K.~Kelly, and N.~Jan.
\newblock Self-avoiding walks in two to five dimensions: exact enumerations and
  series study.
\newblock \emph{J. Phys. A: Math. Gen.}, 25:\penalty0 1429--1440, 1992.

\bibitem[MacDonald et~al.(2000)MacDonald, Joseph, Hunter, Moseley, Jan, and
  Guttmann]{macdonald00}
D.~MacDonald, S.~Joseph, D.~L. Hunter, L.~L. Moseley, N.~Jan, and A.~J.
  Guttmann.
\newblock Self-avoiding walks on the simple cubic lattice.
\newblock \emph{J. Phys. A: Math. Gen.}, 33:\penalty0 5973--5983, 2000.

\bibitem[Clisby et~al.(2007)Clisby, Liang, and Slade]{clisby07}
Nathan Clisby, Richard Liang, and Gordon Slade.
\newblock Self-avoiding walk enumeration via the lace expansion.
\newblock \emph{J. Phys. A: Math. Theor.}, 40:\penalty0 10973--11017, 2007.

\bibitem[Schram et~al.(2011)Schram, Barkema, and Bisseling]{schram11}
R.~D. Schram, G.~T. Barkema, and R.~H. Bisseling.
\newblock Exact enumeration of self-avoiding walks.
\newblock \emph{J. Stat. Mech.}, P06019, 2011.

\bibitem[Butera and Comi(1997)]{Butera1997Nvectorspin}
P.~Butera and M.~Comi.
\newblock $n$-vector spin models on the simple-cubic and the
  body-centered-cubic lattices: A study of the critical behavior of the
  susceptibility and of the correlation length by high-temperature series
  extended to order $\beta^{21}$.
\newblock \emph{Phys. Rev. B}, 56:\penalty0 8212--8240, 1997.

\bibitem[Martin et~al.(1967)Martin, Sykes, and
  Hioe]{Martin1967ProbabilityofInitialRingClosureGorSelfAvoidingWalks}
J.~L. Martin, M.~F. Sykes, and F.~T. Hioe.
\newblock Probability of initial ring closure for self avoiding walks on the
  face centered cubic and triangular lattices.
\newblock \emph{J. Chem. Phys.}, 46:\penalty0 3478--3481, 1967.

\bibitem[McKenzie(1979)]{McKenzie1979SAWsFCC}
S.~McKenzie.
\newblock Self-avoiding walks on the face-centred cubic lattice.
\newblock \emph{J. Phys. A: Math. Gen.}, 12:\penalty0 L267, 1979.

\bibitem[de~Gennes(1979)]{degennes79}
Pierre-Giles de~Gennes.
\newblock \emph{Scaling Concepts in Polymer Physics}.
\newblock Cornell University Press, Ithaca, NY, 1979.

\bibitem[Clisby and D\"unweg(2016)]{Clisby2016HydrodynamicRadiusForSAWs}
Nathan Clisby and Burkhard D\"unweg.
\newblock High-precision estimate of the hydrodynamic radius for self-avoiding
  walks.
\newblock \emph{Phys. Rev. E}, 94:\penalty0 052102, 2016.

\bibitem[Schram et~al.(2013)Schram, Barkema, and Bisseling]{schram13}
Raoul~D. Schram, Gerard~T. Barkema, and Rob~H. Bisseling.
\newblock {SAWdoubler}: a program for counting self-avoiding walks.
\newblock \emph{Comput. Phys. Commun.}, 184:\penalty0 891--898, 2013.

\bibitem[Guttmann(2016)]{guttmann16}
A.~J. Guttmann.
\newblock Series extension: predicting approximate series coefficients from a
  finite number of exact coefficients.
\newblock \emph{J. Phys. A: Math. Theor.}, 49:\penalty0 415002, 2016.

\bibitem[Guttmann(1989{\natexlab{b}})]{guttmann89b}
A.~J. Guttmann.
\newblock \emph{Asymptotic Analysis of Power-Series Expansions}, volume~13 of
  \emph{Phase Transitions and Critical Phenomena}.
\newblock Academic Press, 1989{\natexlab{b}}.

\bibitem[Clisby and Jensen(2012)]{Clisby2012newtransfermatrix}
Nathan Clisby and Iwan Jensen.
\newblock A new transfer-matrix algorithm for exact enumerations: self-avoiding
  polygons on the square lattice.
\newblock \emph{J. Phys. A: Math. Theor.}, 45:\penalty0 115202, 2012.

\bibitem[Jensen(2016)]{Jensen2016SquareLatticeSAWsBiasedDifferentialApproximants}
Iwan Jensen.
\newblock Square lattice self-avoiding walks and biased differential
  approximants.
\newblock \emph{J. Phys. A: Math. Theor.}, 49:\penalty0 424003, 2016.

\bibitem[Clisby(2017)]{Clisby2017Scale-freeGammaSAWsArxiv}
Nathan Clisby.
\newblock Scale-free {M}onte {C}arlo method for calculating the critical
  exponent $\gamma$ of self-avoiding walks, January 2017.
\newblock URL \url{http://arxiv.org/abs/1701.08415}.

\bibitem[Shimada and Hikami(2016)]{Shimada2016SAWsIn3DConformalBootstrap}
Hirohiko Shimada and Shinobu Hikami.
\newblock Fractal dimensions of self-avoiding walks and ising high-temperature
  graphs in 3d conformal bootstrap.
\newblock \emph{J. Stat. Phys.}, 165:\penalty0 1006--1035, 2016.

\bibitem[Clisby(2010)]{Clisby2010AccurateEstimateCritical}
Nathan Clisby.
\newblock Accurate estimate of the critical exponent $\nu$ for self-avoiding
  walks via a fast implementation of the pivot algorithm.
\newblock \emph{Phys. Rev. Lett.}, 104:\penalty0 055702, 2010.

\bibitem[Hsu and Grassberger(2004)]{Hsu2004Polymersconfinedbetween}
Hsiao-Ping Hsu and Peter Grassberger.
\newblock Polymers confined between two parallel plane walls.
\newblock \emph{J. Chem. Phys.}, 120:\penalty0 2034--41, 2004.

\bibitem[Prellberg(2001)]{Prellberg2001Scalingselfavoiding}
T.~Prellberg.
\newblock Scaling of self-avoiding walks and self-avoiding trails in three
  dimensions.
\newblock \emph{J. Phys. A: Math. Gen.}, 34:\penalty0 L599--L602, 2001.

\bibitem[Caracciolo et~al.(1998)Caracciolo, Causo, and
  Pelissetto]{Caracciolo1998Highprecisiondetermination}
Sergio Caracciolo, Maria~Serena Causo, and Andrea Pelissetto.
\newblock High-precision determination of the critical exponent $\gamma$ for
  self-avoiding walks.
\newblock \emph{Phys. Rev. E}, 57:\penalty0 R1215--R1218, 1998.

\bibitem[Guida and Zinn-Justin(1998)]{Guida1998CriticalexponentsN}
R.~Guida and J.~Zinn-Justin.
\newblock Critical exponents of the {$N$}-vector model.
\newblock \emph{J. Phys. A: Math. Gen.}, 31:\penalty0 8103--8121, 1998.

\bibitem[Belohorec(1997)]{Belohorec1997Renormalizationgroupcalculation}
Peter Belohorec.
\newblock \emph{Renormalization group calculation of the universal critical
  exponents of a polymer molecule}.
\newblock PhD thesis, University of Guelph, 1997.

\bibitem[Li et~al.(1995)Li, Madras, and
  Sokal]{Li1995CriticalExponentsHyperscaling}
Bin Li, Neal Madras, and Alan~D. Sokal.
\newblock Critical exponents, hyperscaling, and universal amplitude ratios for
  two- and three-dimensional self-avoiding walks.
\newblock \emph{J. Stat. Phys.}, 80:\penalty0 661--754, 1995.

\end{thebibliography}
\end{document}